\documentstyle[11pt,newpasp,twoside,wrapfig,psfig]{article}
\index{SNR}
\index{Pulsars}
\index{Spectroscopy}

\def\amin{\ifmmode^{\prime}\else$^{\prime}$\fi}
\def\asec{\ifmmode^{\prime\prime}\else$^{\prime\prime}$\fi}

\newcommand\asca{{\sl ASCA}}
\def\nhunits{\times 10^{22}\ {\rm cm^{-2}}}

\def\edcomment#1{\iffalse\marginpar{\raggedright\sl#1\/}\else\relax\fi}
\reversemarginpar

\begin{document}
\title{Chandra X-ray Spectroscopy of Kes~75, its Young Pulsar, and 
its Synchrotron Nebula}
\author{B. F. Collins, E. V. Gotthelf, and D. J. Helfand}
\affil{Columbia Astrophysics Laboratory, 550 West 120th St, New York, NY 10027, USA}

\begin{abstract}
We have observed the young Galactic supernova remnant Kes~75 with the 
Chandra X-ray Observatory. This object is one of an increasing number of 
examples of a shell-type remnant with a central extended radio core 
harboring a pulsar.  Here we present a preliminary spatially resolved 
spectroscopic analysis of the Kes~75 system.  We find that the spectrum of 
the pulsar is significantly harder than that of the wind nebula, and both of 
these components can be isolated from the diffuse thermal emission that 
seems to follow the same distribution as the extended radio shell. When we 
characterize the thermal emission with a model of an under-ionized plasma 
and non-solar elemental abundances, we require a significant diffuse high 
energy component, which we model as a power-law with a photon index similar 
to that of the synchrotron nebula.
\end{abstract}

\section{Introduction}

Kes 75 (also known as G29.7$-$0.3) is one example in our
Galaxy of a young, shell-type remnant ($3.5\arcmin$ in diameter)
with a central core ($30\arcsec$) whose observed properties
suggest a synchrotron nebula similar to the Crab Nebula (Becker,
Helfand \& Szymkowiak 1983; Becker \& Helfand 1984; Blanton \& Helfand
1996).  Observations by the Advanced Satellite for Cosmology and 
Astrophysics (\asca) verified the existence of both thermal and 
non-thermal emission, but lacked the spatial resolution
necessary to separate the components (Blanton \& 
Helfand 1996).  Monitoring with the Rossi X-Ray Timing Explorer led Gotthelf 
et al. (2000) to discover 
a 700 year old pulsar, PSR~J1846$-$0258, located in the \asca\ data within 
the Crab-like core.
Expanding on the previous work, we discover a significant spectral difference 
between the pulsar and the surrounding 
wind nebula.  Using a non-equilibrium ionization 
model we find different temperature and heavy-metal abundances in separate
regions of the shell, as well as evidence of non-thermal emission
throughout.

\section{Observations}

We report here on an observation of the supernova remnant Kes 75
obtained on $10-11$ Oct 2000 with the Chandra X-ray Observatory
(Weisskopf, O'Dell, \& van Speybroeck 1996). Photons were collected
using the Advanced CCD

%\begin{wrapfigure}{L}{2.52in}
\begin{figure}[h]
\begin{minipage}{0.48\linewidth}
\psfig{file=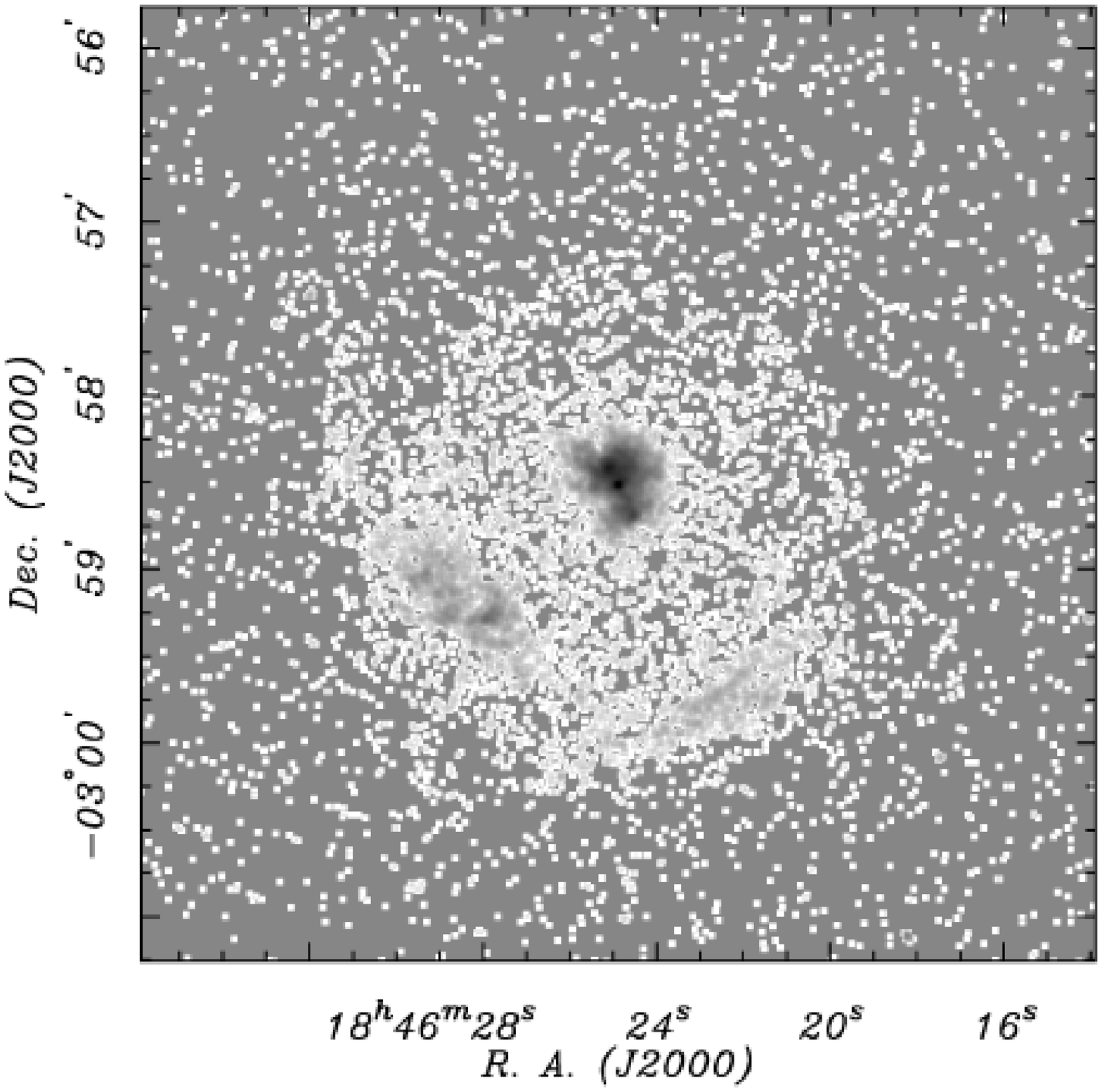,width=\linewidth,angle=-90.0}
\vspace{-.3in}
a)
 \end{minipage}
\begin{minipage}{0.48\linewidth}
\vspace{0.34in}
%\hspace{0.35in}

\hspace{0.35in} 
\psfig{file=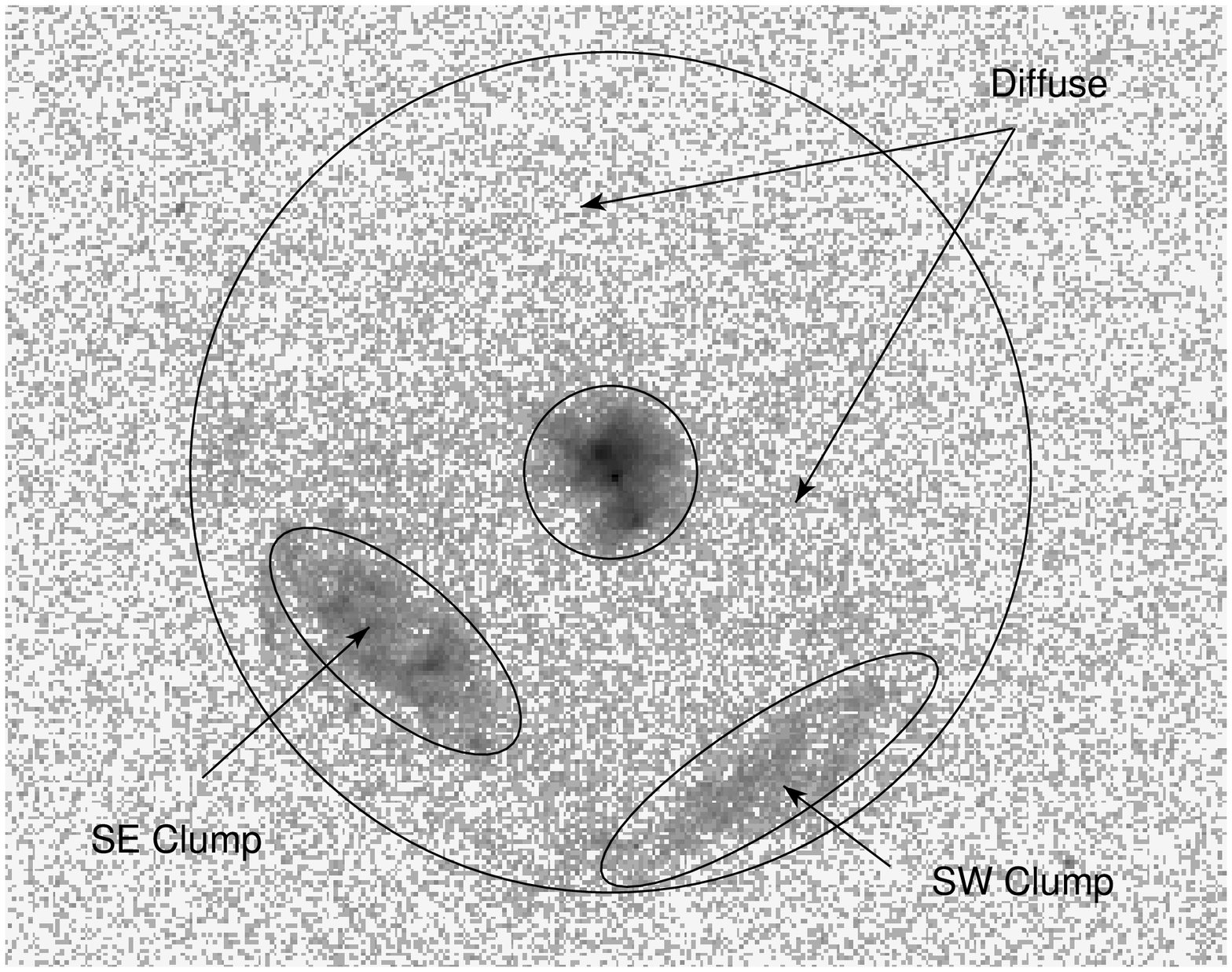,width=1.8in,angle=0,clip=}
\vspace{0.315in}

%\vspace{0.24in} 
%{\bf Figure 2.} \
\end{minipage}

\vspace{0.3in}
{\bf Figure 1.} \small a) Chandra X-ray image of Kes~75.  Logarithmic scaling is used to highlight the smooth, fainter emission and the overall circular symmetry.  The pulsar is indicated by the central bright spot surrounded by the bright wind nebula. b) Kes~75 with spectral extraction regions indicated.
\end{figure}

\noindent Imaging Spectrometer (ACIS), with the target
placed on S3 chip.  This back-side illuminated CCD is sensitive to
photons in the $\sim 0.2-10$ keV energy range with a spectral
resolution of $E/\Delta E \sim 10$ at 1 keV and an on-axis angular
resolution of $\sim 0\farcs5$ at 1 keV.  We processed Level 1 data 
with the charge transfer inefficiency reduction algorithms created by 
Townsley et al. (2000), and then chose the \asca-like 
[0, 2, 3, 4, 6] grade set from the resulting event list for our 
analysis.  We created a lightcurve from the non-source data on the
ACIS-S3 chip to select Good-Time Intervals free of 
unusually high background activity. After all processing,
we had obtained 34 ks of exposure time. A high level of absorption
toward the source blocked most photons with energy $\leq 1$, so 
all models were fit on the energy range of 
$1-7$ keV.  All spectra were grouped to contain a minimum of $20\ 
{\rm counts\ bin^{-1}}$, and all errors are quoted to the 1-$\sigma$
confidence range.

\section{Results}

We centered a small circular
aperture with an $\sim 2\arcsec$ radius on PSR~J1846$-$0258, and a larger 
elliptical region to contain the entire pulsar wind nebula. 
The count rate of the pulsar itself produces noticeable CCD
pile-up effects and renders background contamination insignificant.
For the wind nebula and its subdivisions,
we subtracted background spectra from a patch of smooth emission just 
southwest of the core, but within the shell of the supernova remnant;
the count rate for the nebula is insufficient to cause concern over pile-up.
Because all of the photons from the central source fall onto a 
relatively small area of the CCD, the instrumental response for all of 
the regions can be described sufficiently with the ACIS-S3 RMF provided 
with the CTI reduction software and a single point source mirror response 
(ARF) created for PSR~J1846$-$0258 with the standard CIAO tools, but 
incorporating the supplied CTI-adjusted QEU file.

In the center of the remnant, the pulsar and its wind nebula are both
clearly resolved. Outside of the bright central nebula, 
an expanse of faint diffuse emission
is visible, with two brightened features along the southern edge.  We
encompass all of the diffuse shell with a $200\asec$ wide circular 
aperture, centered
on PSR~J1846$-$0258.  From this region we exclude
the elliptical region of the pulsar and its wind nebula.   
We also drew an ellipse around each of the `clumps' 
in the south, to analyze them separately, and exclude them from the fainter
emission.  For the background subtraction of all regions, we used an 
off-source area of the S3 chip, to the northwest of Kes~75.  Using the 
point-like ARF no longer sufficed for our extended faint source.  Instead, 
we used software created by A. Vikhlinin to weight
each 32$\times$32 pixel response region by X-ray flux, and average them
together over the whole extraction aperture, again incorporating the new QEU 
file.  We continued to use the
supplied CTI-adjusted RMF.

\subsection{PSR~J1846$-$0258 and the Synchrotron Nebula}

To account for the phenomenon of CCD pile-up in our observation, we employed 
the {\tt pileup} model included in XSPEC v11.1.0p.  
Since this model is still in the testing phase, we stress the preliminary
nature of our result.  We fit the spectrum with an 
absorbed power-law, which consistently produced better fits than blackbody or 
bremsstrahlung continua.  The best fit parameters of the piled-up, absorbed
power-law model are $\Gamma = 1.2 \pm 0.1$, and $N_H = 3.5 \pm 0.2 \nhunits$,
with a reduced $\chi^2_\nu$ = 0.86 (58 dof).  The best-fit column density seemed 
to border on agreement with the best-fit column density of the entire wind 
nebula ($3.95 \pm 0.07 \nhunits$). 
Since it is not piled-up and is less affected by the possibly extreme
environments immediately surrounding a pulsar, we decided this was a more 
reliable estimate, froze the parameter, and re-fit.  The photon index 
increased to 1.5 $\pm 0.1$.

Compared to the pulsar, the central elliptical 
region shows a spectra just as devoid of 
emission lines that produces equally poor fits to thermal models.
In terms of the power-law 
photon index, the nebula is unique ($\Gamma = 1.91 \pm 0.04$).  Confidence
contours of the column density and photon index for the entire wind nebula
region demonstrate that the two regions are spectrally distinct.

\subsection{The Thermal Shell}

To analyze the emission from the shell of Kes~75, we followed two approaches.
Using an arbitrary thermal bremsstrahlung continuum with
three Gaussians to represent the most prominent emission lines, 
we aimed to reproduce the thermal components of a model fit to the \asca\ 
observation by Blanton \& Helfand (1996), and to
compare the spectra from different parts of the shell.
In the other case we chose models that more appropriately characterized  
emission from the ionized plasma of a supernova remnant.

Fitting the arbitrary bremsstrahlung and Gaussians model to the data
extracted ~15\arcsec\ to 100\arcsec\ from the pulsar, the 
positions of the three strongest spectral lines correspond to the ions 
Mg~{\small XI}, 
Si~{\small XIII}, and S~{\small XV}, and are all consistent with their 
counterparts
in the \asca\ analysis.  The plasma temperature and column density of
the best-fit model, however, disagree drastically.  

The spectra extracted from the bright clumps in the southern portion of
the remnant, when compared the emission elsewhere in the shell, exhibit 
similar
Mg~{\small XI}, Si~{\small XIII}, and S~{\small XV} lines, and yield the 
same best-fit column density.  The values of 
$kT$ are much higher in the clumps than in the faint background,
indicating only that their spectra are generally softer.  

A single non-equilibrium ionization collisional plasma model 
(Borkowski et al. 2001) with interstellar absorption and varying 
abundances fit both the continuum and the emission lines of each
spectra poorly.
When we add a power-law component, however, the fits improve dramatically.  
We froze
the column density at the value fit to the pulsar wind
nebula, and fit each region twice, once varying the elemental abundances
together in solar ratios, and once letting each element whose X-ray 
emission lies primarily within the energy range of our analysis vary
independently.  The results indicate different elemental ratios and ion 
densities for the different regions of the remnant.  Additional in depth 
analyses of these models will be presented in a forthcoming paper.

\section{Discussion}

For the \asca\ observation of Kes~75, it was not possible
to extract only the spectrum of the central Crab-like component of the 
supernova remnant without
the surrounding thermal part, only to fit a model to all the data that 
accounts
for both of the components.  Given this limitation, Blanton \& Helfand (1996) 
characterized all the non-thermal emission of the object with a 
single power-law ($\Gamma = 2.0$).  We can now 
resolve the core from the shell, and analyze their spectra individually.
The clearly non-thermal spectrum of the central nebula exhibits a
photon index consistent with the \asca\ findings, yet if this were the sole
source of non-thermal flux, our heuristic model of the shell portion should 
be just as consistent with the previous results.  The lack of similarity 
between the two spectra indicates a significant 
amount of non-thermal flux from outside the central core of the remnant.
This extra emission may account for the spectral inconsistencies between 
the line strengths and the thermal continuum.

\acknowledgments
This research is supported by CXC grant SAOGO0-1130X.
E. V. G. is supported by NASA LTSA grant NAG5-22250.

\end{document}